\documentclass{article}
\usepackage{amsmath}
\usepackage[utf8]{inputenc}
\usepackage[margin=1in]{geometry}
\usepackage[T1]{fontenc}
\usepackage{graphicx}
\usepackage{booktabs, bbm}
\usepackage{float}
\usepackage{amssymb}
\usepackage{upquote}
\usepackage{pseudocode}
\usepackage{tikz}
\usepackage{algorithm}
\usepackage{algpseudocode}
\usepackage{parskip}
\usepackage{scrextend}
\usepackage{fancyhdr}
\usepackage{pgfplots}
\usepackage{mathtools}
\usepackage[most]{tcolorbox}
\usepackage{xparse}
\usepackage{bm}

\usetikzlibrary{arrows.meta}

\tikzset{
    myarrow/.style={-{Triangle[length=3mm,width=1mm]}}
}

\definecolor{darkpastelgreen}{rgb}{0.01, 0.75, 0.24}

\newtheorem{innercustomgeneric}{\customgenericname}
\providecommand{\customgenericname}{}
\newcommand{\newcustomtheorem}[2]{%
  \newenvironment{#1}[1]
  {%
   \renewcommand\customgenericname{#2}%
   \renewcommand\theinnercustomgeneric{##1}%
   \innercustomgeneric
  }
  {\endinnercustomgeneric}
}

\newcustomtheorem{ntheorem}{Theorem}
\newcustomtheorem{nlemma}{Lemma}
\newcustomtheorem{nex}{Exercise}

\DeclarePairedDelimiterX{\innerprod}[2]{\langle}{\rangle}{#1, #2}

\newcommand{\norm}[2]{\left\| #1 \right\|_{#2}}

\newcommand{\R}{\mathbb{R}}

\newcommand{\E}{\mathbb{E}}

\newcommand{\mbf}[1]{\mathbf{#1}}
\newcommand{\mbx}{\mathbf{x}}
\newcommand{\mby}{\mathbf{y}}

\DeclareMathOperator*{\argmin}{arg\,min}

\usepackage{spconf}
\usepackage{subcaption}

\usepackage{enumitem}
\setlist{nosep, leftmargin=14pt}

\usepackage{mwe} % to get dummy images

\newcommand\blfootnote[1]{%
  \begingroup
  \renewcommand\thefootnote{}\footnote{#1}%
  \addtocounter{footnote}{-1}%
  \endgroup
}

\title{Scale-Cascaded Diffusion Models for Super-Resolution in Medical Imaging}
\name{Darshan Thaker\textsuperscript{*}, Mahmoud Mostapha$^\dagger$, Radu Miron$^{\dagger \dagger}$, Shihan Qiu$^\dagger$, Mariappan Nadar$^\dagger$}
\address{\textsuperscript{*} University of Pennsylvania, Philadelphia, PA USA \qquad $^\dagger$ Siemens Healthineers, Princeton, NJ USA \\ $^{\dagger \dagger}$ Siemens Industry Software, Romania (Advanta)}
\begin{document}
%\ninept
%
\maketitle
\begin{abstract}
Diffusion models have been increasingly used as strong generative priors for solving inverse problems such as super-resolution in medical imaging. However, these approaches typically utilize a diffusion prior trained at a single scale, ignoring the hierarchical scale structure of image data. In this work, we propose to decompose images into Laplacian pyramid scales and train separate diffusion priors for each frequency band. We then develop an algorithm to perform super-resolution that utilizes these priors to progressively refine reconstructions across different scales. Evaluated on brain, knee, and prostate MRI data, our approach both improves perceptual quality over baselines and reduces inference time through smaller coarse-scale networks. Our framework unifies multiscale reconstruction and diffusion priors for medical image super-resolution.
\end{abstract}
\begin{keywords}
Super-resolution, Diffusion models, Posterior Sampling, Scale-cascaded
\end{keywords}
\section{Introduction}
\label{sec:intro}

\blfootnote{Work performed while Darshan was an intern at Siemens Healthineers. The concepts and information presented in this paper/presentation are based on research results that are not commercially available. Future commercial availability cannot be guaranteed.}

The task of super-resolution deals with reconstructing a high-resolution image $\mbx$ from a low-resolution measurement $\mby$ obtained via
\begin{equation}
    \mby = \mbf{H} \mbx + \mbf{z}
\end{equation}
where $\mbf{H}$ denotes a downsampling operator and $\mbf{z}$ is additive Gaussian noise. Because there is a strong loss of information through the non-invertible operator $\mbf{H}$, strong priors are needed for $\mbx$ to ensure high-quality reconstruction \cite{rudin1992nonlinear, bora2017compressed}. A popular Bayesian approach, known as posterior sampling, is to develop a conditional sampler for $p(\mbx \mid \mby)$ using a powerful unconditional generative prior for $p(\mbx)$ such as a diffusion model \cite{zhu2023denoising, chung2022diffusion, daras2024survey}. A key advantage of this approach is that it does not require retraining the diffusion prior under different choices of measurement. For diffusion models, posterior sampling is performed by alternating unconditional denoising steps and data consistency steps that enforce alignment of the diffusion iterates with the measurement $\mby$.

%Diffusion models generate samples from the data distribution by learning the score function of the data distribution under different noise levels. For the conditional sampling task, this score function $\nabla_{\mbx_t} \log p(\mbx_t \mid \mby$ decomposes as $\nabla_{\mbx_t} \log p(\mbx)t) + \nabla_{\mbx_t} \log p(\mby \mid \mbx_t)$ where $\mbx_t$.

While diffusion-based posterior samplers have achieved strong results, most operate at a single image scale. Instead, recent work has recognized that diffusion processes naturally follow a coarse to fine generation structure, which has led to the development of posterior sampling algorithms that exploit this emergent behavior during inference \cite{thaker2025frequency, rissanen2022generative}. However, these approaches still rely on a single diffusion prior trained at a fixed scale.

In this work, we directly embed scale-space structure into the unconditional diffusion prior itself. To achieve this, we exploit the Laplacian pyramid representation of images \cite{burt1987laplacian}. In the Laplacian pyramid, an image is decomposed into multiple scales where each scale representation captures a certain frequency band. This allows us to train separate diffusion models at each scale, each specialized to a specific frequency range. Further, we take inspiration from classical multigrid approaches for solving inverse problems in order to develop a scale-cascaded posterior sampling algorithm that exploits the scale-space decomposition \cite{laurent2025multilevel, oh2003multigrid}. In this algorithm, a given super-resolution task is decomposed into a sequence of smaller, better-conditioned super-resolution tasks, and coarser scale reconstructions are used to better initialize finer-scale samplers. Our main contributions are as follows:

\begin{enumerate}
    \item We construct a scale-cascaded diffusion prior defined over Laplacian pyramid scales and train this prior on a large dataset of MR images of different anatomies. 
    \item Given this prior, we introduce a scale-cascaded posterior sampling algorithm that decomposes a given super-resolution task into a sequence of smaller and better-conditioned super-resolution tasks. In this sequence, coarser scale reconstructions are used to better initialize finer-scale samplers.
    \item We evaluate our algorithm on various super-resolution tasks at different downsampling factors. We observe that our algorithm improves the quality of reconstructions compared to single-scale posterior sampling algorithms. Further, our formulation reduces computational cost by using smaller diffusion networks at coarse scales. 
\end{enumerate}

Together, this framework unifies classical multiscale reconstruction with modern diffusion-based priors, providing a principled and efficient approach for super-resolution.

\section{Background}
\label{sec:background}

Before providing the details of our method, we first provide some background on diffusion models and posterior sampling. Diffusion models are generative models that learn to reverse a gradual noising process \cite{ho2020denoising}. This noising process is constructed over the interval $t \in [0, 1]$ such that samples at $t = 0$ are clean data and samples at $t = 1$ are isotropic Gaussian. Sampling from a data distribution proceeds by starting with a sample $\mbx_t$ at $t = 1$, progressively denoising through estimates of $E[\mbx_0 \mid \mbx_t]$ and adding back noise. Given a pretrained unconditional diffusion prior, posterior sampling algorithms such as DiffPIR further introduce data consistency steps in addition to the denoising steps \cite{zhu2023denoising}. Given $\hat{\mbx}_0 \triangleq \E[\mbx_0 \mid \mbx_t]$,  the data consistency steps estimate $E[\mbx_0 \mid \mbx_t, \mby]$ as the solution to an optimization problem
\begin{equation} \label{eq:diffpir}
	\argmin_{\mbx} \underbrace{\norm{\mby - \mbf{H} \mbx}{2}^2}_{\text{Data Fitting}} + \underbrace{\tau \norm{\mbx - \hat{\mbx}_0}{2}^2}_{\text{Prior}}
\end{equation}
 For more details, we refer the reader to \cite{zhu2023denoising, daras2024survey}. 

\section{Method}
\label{sec:method}

\subsection{Laplacian Pyramid Construction}

We first describe the Laplacian pyramid decomposition we utilize in our method. For a image $\mbx \in \R^{d \times d}$, a 3-level Laplacian pyramid yields a hierarchy $\{\mbx^{(3)}, \mbx^{(2)}, \mbx^{(1)}\}$ where each $\mbx^{(i)} \in \R^{\frac{d}{2^{i - 1}} \times \frac{d}{2^{i - 1}}}$ such that
\begin{align}
	\mbx^{(3)} &= down(down(\mbx)) \\
	\mbx^{(2)} &= down(\mbx) - up(\mbx^{(3)}) \\
	\mbx^{(1)} &= \mbx - up(down(\mbx)).
\end{align}
Above, $down$ and $up$ represent downsampling and upsampling operations with factor 2. Each image represents a frequency band with $\mbx^{(3)}$ as the coarsest representation of the image and subsequent scales representing difference images between successive pyramid levels. For complex-valued images, we apply the pyramid on the real and imaginary components of the image independently. 

\subsection{Scale-Cascaded Diffusion Prior}

We train three diffusion priors at each scale of the Laplacian pyramid that model $p(\mbx^{(3)}), p(\mbx^{(2)} \mid \mbx^{(3)}),$ and $p(\mbx^{(1)} \mid \mbx^{(2)}, \mbx^{(3)})$. Specifically, the coarsest scale representation, which we refer to as our level 3 model, is an unconditional diffusion model, whereas the finer scale diffusion models (level 2 and level 1) are conditioned on the outputs of the coarser scale diffusion models. The conditioning is implemented as upsampling the coarser scale representations and concatenating channel-wise to the model input. At the $x^{(1)}$ scale, we condition the diffusion model on $\mbx^{(2)} + up(\mbx^{(3)}) $ instead of $\mbx^{(2)}$ and $\mbx^{(3)}$ separately. In contrast to prior scale-cascaded diffusion models \cite{ho2022cascaded}, our model explicitly predicts the difference image representations at finer scales, reducing the training task to a certain frequency band \cite{lai2017deep, denton2015deep}. 

\subsection{Scale-Cascaded Posterior Sampling}

Given our scale-cascaded diffusion prior at 3 levels, we develop a scale-cascaded posterior sampling algorithm for super-resolution tasks. Consider a super-resolution task where $\mbf{H}$ represents a downsampling operator of factor $k$. Our algorithm sequentially applies the DiffPIR algorithm from level 3 to level 1. Specifically, at each level $i$, we replace the data fitting term in \eqref{eq:diffpir} with 
\begin{equation}
	\norm{y - \mbf{H}_2 \left(\mbx^{(i)} + \sum_{j = i + 1}^{3} up(\mbx^{(j)}) \right)}{2}^2
\end{equation} 
and the prior term is applied with the estimate at level $i$. Above, the upsampling operation $up$ matches the resolution of $\mbx^{(i)}$ and $\mbx^{(j)}$. Further, we decompose $\mbf{H}$ into a sequence of 2x downsampling operators, denoted as $\mbf{H}_2$. This ensures that data consistency is applied in a coarse-to-fine manner as we sample from the three diffusion models sequentially. Thus, the problem remains better-conditioned and the coarse scale outputs provide valuable conditioning information for the following levels. Compared to multi-grid methods \cite{oh2003multigrid}, we perform posterior sampling at the scale of each diffusion model by modifying the forward operator as opposed to upsampling $\mbx^{(i)}$ and retaining the same $\mbf{H}$. 

\begin{table*}[t] 
\small
\centering
\begin{tabular}{lccc ccc ccc}
\toprule
& \multicolumn{3}{c}{\textbf{Brain}} & \multicolumn{3}{c}{\textbf{Knee}} & \multicolumn{3}{c}{\textbf{Prostate}} \\
\cmidrule(lr){2-4} \cmidrule(lr){5-7} \cmidrule(lr){8-10}
\textbf{Method} & PSNR$\uparrow$ & SSIM$\uparrow$ & LPIPS$\downarrow$
               & PSNR$\uparrow$ & SSIM$\uparrow$ & LPIPS$\downarrow$
               & PSNR$\uparrow$ & SSIM$\uparrow$ & LPIPS$\downarrow$ \\
\midrule
DiffPIR \cite{zhu2023denoising} & 30.14 & 0.95 & 0.11 & 30.57 & 0.86 & 0.15 & 34.44 & 0.94 & 0.11 \\
DPS \cite{chung2022diffusion}   & 26.93 & 0.92 & 0.10 & 27.66 & 0.85 & 0.18 & 30.38 & 0.91 & 0.15 \\
Scale-Cascaded DiffPIR (Ours) & \textbf{31.27} & \textbf{0.96} & \textbf{0.09} 
                       & \textbf{31.63} & \textbf{0.91} & \textbf{0.13} 
                       & \textbf{35.91} & \textbf{0.97} & \textbf{0.08} \\
\bottomrule
\end{tabular}
\caption{\textbf{Quantitative Results on 2x super-resolution.} We find that our approach strongly improves upon posterior sampling baselines in perceptual quality of reconstructions. }
\label{tab:2xsr}
\end{table*}

\begin{table*}[t]
\small
\centering
\begin{tabular}{lccc ccc ccc}
\toprule
& \multicolumn{3}{c}{\textbf{Brain}} & \multicolumn{3}{c}{\textbf{Knee}} & \multicolumn{3}{c}{\textbf{Prostate}} \\
\cmidrule(lr){2-4} \cmidrule(lr){5-7} \cmidrule(lr){8-10}
\textbf{Method} & PSNR$\uparrow$ & SSIM$\uparrow$ & LPIPS$\downarrow$
               & PSNR$\uparrow$ & SSIM$\uparrow$ & LPIPS$\downarrow$
               & PSNR$\uparrow$ & SSIM$\uparrow$ & LPIPS$\downarrow$ \\
\midrule
DiffPIR \cite{zhu2023denoising} & 24.25 & 0.81 & 0.23 & 25.71 & 0.71 & 0.33 & 28.23 & 0.79 & 0.30 \\
DPS \cite{chung2022diffusion} & 22.59 & 0.77 & 0.21 & 23.28 & 0.67 & 0.36 & 25.07 & 0.73 & 0.33 \\
Multi-Grid PnP \cite{laurent2025multilevel} & 23.48 & 0.78 & 0.24 & 24.49 & 0.67 & 0.34 & 26.60 & 0.73 & 0.32 \\
Scale-Cascaded DiffPIR 2-Level (Ours) & 26.53 & 0.87 & 0.19  & 26.53 & 0.75 & 0.28 & \textbf{29.02} & 0.82 & 0.26 \\
Scale-Cascaded DiffPIR 3-Level (Ours) & \textbf{31.69} & \textbf{0.94} & \textbf{0.16} & \textbf{30.76} & \textbf{0.93} & \textbf{0.17} & 28.98 & \textbf{0.90} & \textbf{0.17} \\
\bottomrule
\end{tabular}
\caption{\textbf{Quantitative Results on 4x super-resolution.} Our scale-cascaded approach illustrates the benefit of conditioning information as well as the benefit of multiscale structure in solving inverse problems, improving upon standard posterior sampling algorithms and a standard multi-grid approach.}
\label{tab:4xsr}
\end{table*}

\section{Results}

\subsection{Experimental Setup}

We use the EDM2 backbone for our diffusion priors at different levels \cite{karras2024analyzing}. Our models are trained on the NYU FastMRI dataset consisting of Brain, Knee, and Prostate MRI images from 1.5T, 3T and other scanners \cite{knoll2020fastmri, tibrewala2024fastmri, zbontar2018fastmri} . We evaluate on a held-out dataset of 100 slices per anatomy type and we use the Peak Signal-to-Noise Ratio (PSNR) and Learned Perceptual Image Patch Similarity (LPIPS) metrics. We evaluate our method on two axes. The first isolates the performance of one level of the diffusion model, for which we test a 2x super-resolution task using the level 1 diffusion model. The second evaluates the effect of cascading in posterior sampling, for which we test a 4x super-resolution task. We compare our method to the DiffPIR and DPS algorithms with diffusion models that are not cascaded \cite{zhu2023denoising, chung2022diffusion}. To ensure fair comparison, all models were sampled using $T=200$ steps total, where our cascaded sampling is split into $T/N_{\text{levels}}$ steps at each level. 

\subsection{2x Super-Resolution} 

\begin{figure}[t]
    \centering
    \begin{subfigure}[t]{0.32\linewidth}
        \centering
        \includegraphics[width=2.6cm, height=2.7cm]{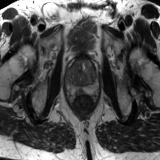}
         \includegraphics[width=2.6cm, height=2.7cm]{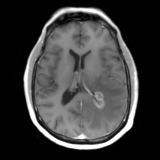}
        \subcaption*{Degraded $\mby$}
    \end{subfigure}\hfill
    \begin{subfigure}[t]{0.32\linewidth}
        \centering
        \includegraphics[width=2.6cm, height=2.7cm]{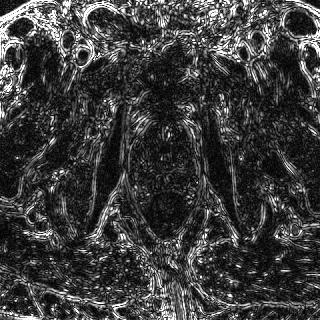}
        \includegraphics[width=2.6cm, height=2.7cm]{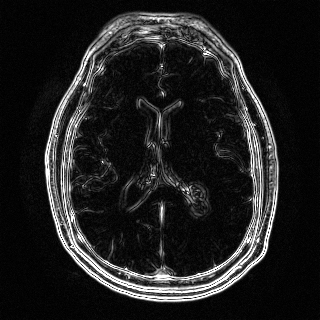}
        \subcaption*{Predicted $\mbx^{(1)}$}
    \end{subfigure}\hfill
    \begin{subfigure}[t]{0.32\linewidth}
        \centering
        \includegraphics[width=2.6cm, height=2.7cm]{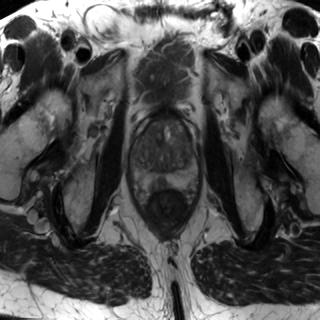}
        \includegraphics[width=2.6cm, height=2.7cm]{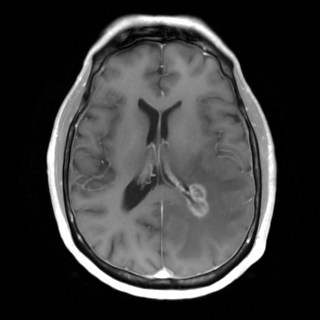}
        \subcaption*{Reconstructed image}
    \end{subfigure}
    \caption{\textbf{Qualitative Results on 2x super-resolution.} Across different anatomies, our algorithm is able to fill in fine-grained detail of the degraded image as shown in the difference image prediction $\mbx^{(1)}$.
    }
    \label{fig:srx2}
\end{figure}

Table \ref{tab:2xsr} shows our results on the 2x super-resolution task using only the level 1 diffusion model and the low-resolution image as the condition. Our method is able to improve the perceptual quality of reconstructions compared to baseline methods, indicating the benefit of introducing conditioning into the diffusion model. This is further illustrated qualitatively in Figure \ref{fig:srx2}. Further, because our approach only predicts the difference image, a smaller model suffices to model the denoising network.

\subsection{4x Super-Resolution}

\begin{figure}[t]
    \centering
    \begin{subfigure}[t]{0.11\textwidth}
        \centering
        \includegraphics[width=2cm, height=4cm]{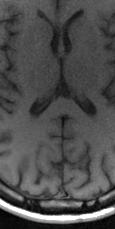}
        \subcaption*{Clean $\mbx$}
    \end{subfigure}\hfill
    \begin{subfigure}[t]{0.11\textwidth}
        \centering
        \includegraphics[width=2cm, height=4cm]{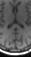}
        \subcaption*{Degraded $\mby$ }
    \end{subfigure}\hfill
    \begin{subfigure}[t]{0.11\textwidth}
        \centering
        \includegraphics[width=2cm, height=4cm]{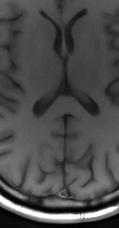}
        \subcaption*{DiffPIR}
    \end{subfigure}\hfill
    \begin{subfigure}[t]{0.11\textwidth}
        \centering
        \includegraphics[width=2cm, height=4cm]{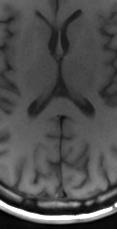}
        \subcaption*{Ours}
    \end{subfigure}
    \caption{\textbf{Qualitative Results on 4x super-resolution.} Compared to DiffPIR, our scale-cascaded algorithm better captures finer anatomy details, resulting in higher quality reconstructions. Zoomed in for clarity.}
    \label{fig:srx4}
\end{figure}

Table \ref{tab:4xsr} and Figure \ref{fig:srx4} show the performance of our method on the 4x super-resolution task. Our scale-cascaded approach significantly improves upon baseline methods due to better-conditioned inverse problems at each scale and better initialization of finer-scale sampling. To disentangle the effect of these two improvements, we also test a 2-level scale-cascaded model such that the second level is unconditional. Even though this model is unconditional at the coarser scale, on the Brain and Knee datasets we still observe a $\approx 2$ dB increase in PSNR compared to DiffPIR, which confirms our hypothesis that the inverse problem is better-conditioned at the coarser scale. We draw a similar conclusion when we compare our approach to the multi-grid baseline, which keeps the same forward operator at each scale. Lastly, as coarser scales operate at a lower resolution, sampling is faster and we observe a 35\% increase in sampling speed. This demonstrates that our scale-cascaded approach helps not only in improving sampling quality but also reducing inference time.

% Full-width figure (4 images: 006–009)

%\section{Conclusion}

\bibliographystyle{IEEEbib}
\bibliography{strings,refs}

\end{document}